\documentclass[prd, aps, 10pt, twocolumn, nofootinbib, superscriptaddress, preprintnumbers, balancelastpage, longbibliography, floatfix]{revtex4-2}
\usepackage[english]{babel}
\pdfoutput=1 

\usepackage[normalem]{ulem}      
\usepackage{bm}
\usepackage[dvipsnames]{xcolor}
\usepackage{graphicx}
\usepackage[T1]{fontenc} 
\usepackage{bigints}
\usepackage{amsmath, amssymb, mathtools, physics, xfrac, graphicx, subfigure, rotating, multirow, tabularx, booktabs, theorem, moreverb, euscript, psfrag, slashed, mathtools, makecell, adjustbox, dcolumn, bm, graphics, hyperref, lettrine, aas_macros, color, tikz, soul, float}
\usepackage[utf8]{inputenc}

\def\lsim{\mathrel{\raise.3ex\hbox{$<$\kern-.75em\lower1ex\hbox{$\sim$}}}}
\def\gsim{\mathrel{\raise.3ex\hbox{$>$\kern-.75em\lower1ex\hbox{$\sim$}}}}

\newcommand{\be}{\begin{equation}}
\newcommand{\ee}{\end{equation}}
\newcommand{\beq}{\begin{equation}}
\newcommand{\eeq}{\end{equation}}
\usepackage{bm}

\usepackage{soul}

\begin{document}

\title{\boldmath The Impact of Muon and Pion Cooling on the Neutrino Spectrum of NGC 1068}

\author{Carlos Blanco}
\thanks{\href{mailto:carlosblanco2718@princeton.edu}{carlosblanco2718@princeton.edu}}
\affiliation{Pennsylvania State University, Institute for Gravitation and the Cosmos}
\affiliation{Princeton University, Department of Physics}
\affiliation{Stockholm University and The Oskar Klein Centre for Cosmoparticle Physics}

\author{Dan Hooper}
\thanks{\href{mailto:dwhooper@wisc.edu}{dwhooper@wisc.edu}}
\affiliation{University of Wisconsin-Madison, Department of Physics and the Wisconsin IceCube Particle Astrophysics Center}

\author{Tim Linden}
\thanks{\href{mailto:linden@fysik.su.se}{linden@fysik.su.se}}
\affiliation{Stockholm University and The Oskar Klein Centre for Cosmoparticle Physics}
\affiliation{Erlangen Centre for Astroparticle Physics (ECAP), Friedrich-Alexander-Universität}

\author{Elena Pinetti}
\thanks{\href{epinetti@flatironinstitute.org}{epinetti@flatironinstitute.org}}
\affiliation{Flatiron Institute, Center for Computational Astrophysics}

\date{\today}

\begin{abstract}

The IceCube Neutrino Observatory has detected a flux of $\sim 1-10 \, {\rm TeV}$ neutrinos from the active galaxy, NGC 1068. The soft spectral index of these neutrinos has previously been interpreted as an indication that this source accelerates protons only up to energies of several hundred TeV. Here, we propose that this source might instead accelerate protons to significantly higher energies, but that the charged pions and muons produced in their interactions undergo significant synchrotron energy losses before they can decay, leading to a cutoff in the neutrino spectrum at TeV-scale energies. This scenario would require very strong magnetic fields to be present in the acceleration region of NGC 1068, on the order of $B \sim 10^7 \, {\rm G}$. We point out that this synchrotron cooling would impact the flavor ratios of the neutrinos from this source, providing a means to test this scenario with future very-large volume neutrino telescopes.  

\end{abstract}

\maketitle
\flushbottom

\section{Introduction}

The IceCube Collaboration has reported the detection of a flux of $\sim 1-10 \, {\rm TeV}$ neutrinos from the nearby ($d=12.7 \, {\rm Mpc}$~\cite{ricci2017bat}) and X-ray bright ($L_X= 7 \times 10^{43} \, {\rm erg/s}$~\cite{Marinucci:2015fqo,Bauer:2014rla}) active galaxy, NGC 1068, with a statistical significance of $4.2\sigma$~\cite{IceCube:2022der}. In the same energy range, this source does not produce a significant flux of gamma rays. In particular, the MAGIC telescope has not detected gamma rays from this source and has placed stringent constraints on its TeV-scale gamma-ray emission~\cite{MAGIC:2019fvw}. Fermi has detected gamma-ray emission from this source, but at much lower energies, $0.1-30 \, {\rm GeV}$~\cite{Blanco:2023dfp,Fermi-LAT:2019yla,Fermi-LAT:2019pir}. 

The lack of TeV-scale gamma rays from NGC 1068 forces us to conclude that this source must produce its neutrinos through pion decay in the dense and optically-thick region that surrounds its supermassive black hole~\cite{Blanco:2023dfp,Das:2024vug,Ajello:2023hkh,Murase:2022dog} (for related works, see Refs.~\cite{Saurenhaus:2025ysu,Ambrosone:2024zrf,IceCube:2024ayt,Inoue:2024nap,Fang:2023vdg}). A growing body of evidence suggests that NGC 1068 is not an outlier in this sense, but rather that a significant fraction of IceCube's diffuse neutrino flux~\cite{IceCube:2020acn,IceCube:2021rpz,IceCube:2013cdw,IceCube:2013low,IceCube:2014stg} originates from sources that are optically thick to gamma rays~\cite{Hooper:2016jls,Murase:2015xka,Giacinti:2015pya,Murase:2013rfa,IceCube:2023htm,IceCube:2019cia,IceCube:2016tpw,IceCube:2018omy,IceCube:2016ipa,Smith:2020oac,IceCube:2016qvd,Hooper:2018wyk}. The dense cores of X-ray-bright active galaxies are particularly interesting in this context~\cite{Murase:2015xka,Khiali:2015tfa,Stecker:2013fxa,Kimura:2014jba,Kalashev:2014vya}.

The neutrino spectrum from NGC 1068, as measured by IceCube, is very soft in its observed energy range, \mbox{$\sim 1-10 \, {\rm TeV}$}, featuring a spectral index of \mbox{$\gamma \sim 3.2 \pm 0.2$~\cite{IceCube:2022der}}. This rapidly falling spectrum could be interpreted as evidence that this source only accelerates protons up to several hundred TeV, leading to a cutoff in the neutrino spectrum at around $\sim 1-10 \, {\rm TeV}$.



In this study, we consider the alternative possibility that NGC 1068 accelerates protons to much higher energies, \mbox{$E_p \gg {\rm PeV}$}, and that the soft neutrino spectrum observed from this source is instead the consequence of the synchrotron cooling that is experienced by charged pions and muons before they decay. Such a scenario benefits from its lack of dependence on a specific proton spectral cutoff, but instead requires extremely strong magnetic fields in the proton acceleration region. Specifically, for an injected proton spectrum of the form $dN_p/dE_p \propto E^{-2}_p$ and magnetic field strengths on the order of $B\sim 10^7 \, {\rm G}$, we find that the predicted neutrino spectrum is in good agreement with that measured by IceCube. This scenario could potentially be tested by measuring the flavor ratios of the neutrinos from this source, but would require a significantly larger detector than IceCube or KM3NeT.

\section{Pion and Muon Cooling in NGC 1068}

In this study, we take the spectrum of the protons accelerated within the corona of NGC 1068 to be characterized by a power-law with an exponential cut-off, 
\begin{align}
\frac{dN_p}{dE_p} \propto E_p^{-\Gamma_p} \, e^{-E_p/E_p^{\rm max}}.
\end{align}
We can estimate the maximum energy of the accelerated protons by comparing the timescales for acceleration, escape (via diffusion), and synchrotron losses~\cite{Dermer:1995ju,Dermer:2014vaa,Kimura:2014jba,Murase:2011cx,Stawarz:2008sp,Murase:2019vdl,Blanco:2023dfp}. The first two of these timescales are given by
\begin{align}
\label{eq:acc}
t_{\rm acc} &=  \dfrac{R \eta c}{\sqrt{3} \, v^2_a} \left(\dfrac{\sqrt{3} \, E_p}{e B R} \right)^{2-q} \\
=&\dfrac{4 \pi \eta c m_p  \tau_T}{B^2 \, \sigma_T } \left(\dfrac{\sqrt{3} \, E_p}{e B R} \right)^{1/3} \nonumber \\
 \approx & \, 0.04 \, {\rm s} \, \bigg(\frac{\eta}{25}\bigg) \bigg(\frac{\tau_T}{0.5}\bigg) \bigg(\frac{25 R_s}{R}\bigg)^{1/3} \bigg(\frac{{\rm MG}}{B}\bigg)^{7/3} \bigg(\frac{E_p}{{\rm PeV}}\bigg)^{1/3} , \nonumber
\end{align}
\begin{align}
\label{eq:diff}
t_{\rm diff} &= \dfrac{3\sqrt{3} \, R}{\eta c} \left(\dfrac{e B R}{\sqrt{3} \, E_p} \right)^{2-q},\\
 \approx & \, 3 \times 10^5 \, {\rm s} \,\bigg(\frac{25}{\eta}\bigg)\bigg(\frac{R}{25 R_s}\bigg)^{4/3}  \bigg(\frac{B}{{\rm MG}}\bigg)^{1/3} \bigg(\frac{{\rm PeV}}{E_p}\bigg)^{1/3}, \nonumber
\end{align}
where $R$ is the radius of the corona, $B$ is the strength of the magnetic field in the corona, and $\eta$ is related to the power spectrum of turbulence, $\eta = [8 \pi \int dk P_k/B^2]^{-1}$, where $P_k \propto k^{-q}$~\cite{Murase:2019vdl}. The quantity, $q$, is the spectral index in momentum space of the particles accelerated by a stochastic magnetic field; here we adopt $q=5/3$, corresponding to the case of Kolmogorov diffusion. The Alfvén velocity is given by $v_A = B/\sqrt{4 \pi m_p n_p}$, where the nucleon density in the corona is $n_p = \sqrt{3} \,\tau_T / \left(\sigma_T R \right)$, and where $\tau_T$ and $\sigma_T$ are the Thomson optical depth and Thomson cross section, respectively. Note that the radius of NGC 1068's corona is estimated to be $R \sim (3-100) \, R_s$~\cite{Murase:2019vdl}, where $R_s = 2 GM_{\rm BH}/c^2$ is the Schwartzchild radius of the system's supermassive black hole. We have taken $M_{
\rm BH} = 2 \times 10^7 \, M_{\odot}$. In producing our numerical results, we will adopt $R=25 \, R_s$, $\eta=25$, and $\tau_T=0.5$, consistent with our previous study~\cite{Blanco:2023dfp}.

Comparing Eqs.~\ref{eq:acc} and~\ref{eq:diff}, we find that for $B \sim {\rm kG}$, one predicts that protons will be accelerated to $E^{\rm max}_p \sim 20 \, {\rm TeV}$ before escaping from the acceleration region. For larger magnetic fields, however, acceleration is more efficient and is instead limited by the rate of synchrotron losses. After averaging over the pitch angle, the timescale for energy losses due to the emission of synchrotron radiation is given by~\cite{Blumenthal:1970gc}
\begin{align}
\tau_{\rm syn} &\equiv \bigg(-\frac{1}{E}\frac{dE}{dt}\bigg)^{-1} = \frac{9 \pi \mu_0 m^4}{e^4  E B^2 v^2}  \\
\approx& \; 4.4 \, {\rm s} \, \bigg(\frac{{\rm PeV}}{E}\bigg) \bigg(\frac{{\rm MG}}{B}\bigg)^2 \bigg(\frac{m}{m_{p}}\bigg)^4. \nonumber
\end{align}

Equating the timescales for acceleration and synchrotron losses as given above, we obtain the following maximum energy to which protons are expected to be accelerated:
\begin{align}
\label{eq:Emax}
E^{\rm max}_p \approx 30 \, {\rm PeV} \, \bigg(\frac{B}{{\rm MG}}\bigg)^{1/4} \, \bigg(\frac{R}{25 \, R_s}\bigg)^{1/4} \, \bigg(\frac{25}{\eta}\bigg)^{3/4} \, \bigg(\frac{0.5}{\tau_T}\bigg)^{3/4}.
\end{align}

In what follows, we take the neutrinos from NGC 1068 to be produced in proton-proton collisions, leading to charged pions which decay into neutrinos: $\pi^+ \rightarrow \nu_{\mu} \mu^+  \rightarrow \nu_{\mu} e^+ \nu_e \bar{\nu}_{\mu}$ and $\pi^- \rightarrow \bar{\nu}_{\mu} \mu^-  \rightarrow \bar{\nu}_{\mu} e^- \bar{\nu}_e \nu_{\mu}$.
In calculating the spectrum of these neutrinos, we make use of the parameterizations provided in Ref.~\cite{Kelner:2006tc}.

The neutrino spectrum of NGC 1068, as reported by the IceCube Collaboration, features a very soft spectral index, $dN_{\nu}/dE_{\nu} \propto E^{-\gamma}$, with $\gamma \sim 3.2 \pm 0.2$~\cite{IceCube:2022der}. As we demonstrated in our previous work~\cite{Blanco:2023dfp}, this spectrum can be fit by adopting parameters corresponding to $\Gamma_p \sim 2.0-2.2$ and $E^{\rm max}_p \sim (2-6) \times 10^5 \ {\rm GeV}$. For these values of $E^{\rm max}_p$, the resulting neutrino spectrum is very soft over the range of energies measured by IceCube (see also, Refs.~\cite{Murase:2019vdl,Murase:2022dog,Fang:2022trf}). Alternatively, we will argue here that it is possible that NGC 1068 might accelerate protons to significantly higher energies, but that charged pions and muons lose much of their energy through synchrotron before they decay, thereby softening the resulting spectrum of neutrinos.

Comparing the timescale for synchrotron losses to the Lorentz boosted lifetime of a particle, we obtain the following result for charged pions,
\begin{align}
\frac{\tau_{\pi^{\pm}}\, \gamma_{\pi^{\pm}}}{\tau_{\rm syn}} \sim \bigg(\frac{E_{\pi^{\pm}}}{110\,{\rm TeV}}\bigg)^2\bigg(\frac{B}{{\rm MG}}\bigg)^{2}, 
\end{align}
and for muons,
\begin{align}
\frac{\tau_{\mu} \, \gamma_{\mu}}{\tau_{\rm syn}} \sim \bigg(\frac{E_{\mu}}{5.9\,{\rm TeV}}\bigg)^2\bigg(\frac{B}{{\rm MG}}\bigg)^{2},
\end{align}
where $\tau_{\pi^{\pm}} = 2.6 \times 10^{-8} \, {\rm s}$ and $\tau_{\mu} = 2.2\times 10^{-6} \, {\rm s}$. From these comparisons, we see that for $B \gsim {\rm MG}$, very high-energy pions and muons will experience significant energy losses before decaying, impacting the resulting spectrum of neutrinos observed by IceCube.

\begin{figure}[t]
\centering
\includegraphics[width=0.98\columnwidth]{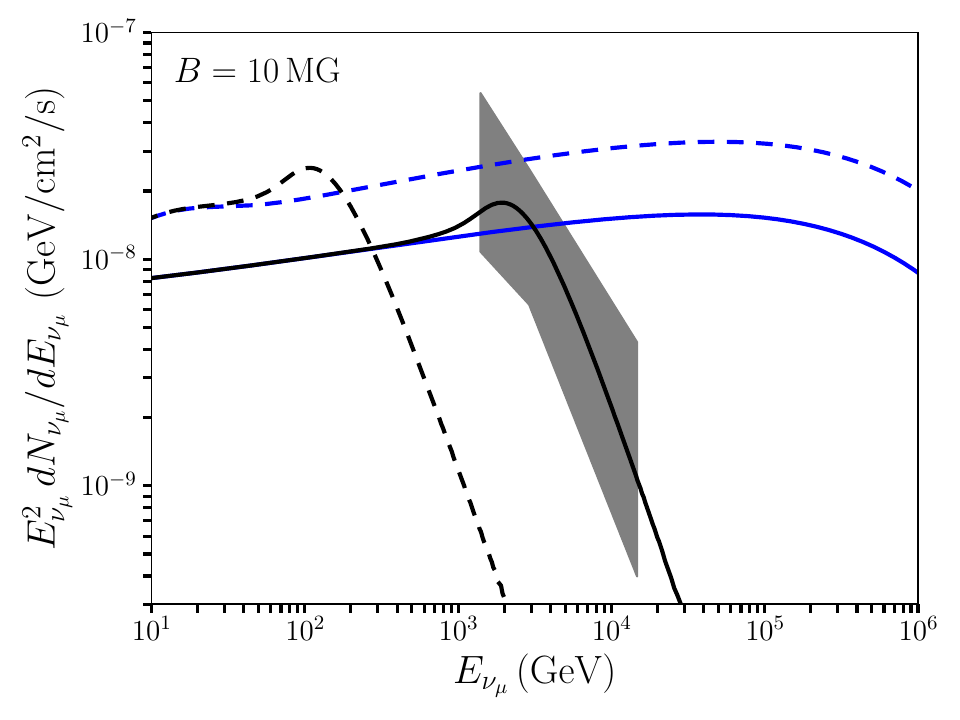}
\caption{The spectrum of muon neutrinos and antineutrinos (after accounting for oscillations) predicted from pion decay ($\pi \rightarrow \mu \nu$, solid curves) and subsequent muon decay ($\mu \rightarrow e \nu \nu$, dashed curves). We have adopted an injected spectral index of $\Gamma_p=2$ and a high-scale energy cutoff, $E^{\rm max}_p \approx 60 \,{\rm PeV}$ (following from Eq.~\ref{eq:Emax}). The blue curves depict the spectrum obtained if pions and muons do not experience significant energy losses due to synchrotron. The black curves, in contrast, include synchrotron losses, adopting a large value of the magnetic field strength, $B=10 \, {\rm MG}$. The grey band represents the spectrum of NGC 1068 as measured by the IceCube Collaboration~\cite{IceCube:2022der}.}
\label{fig:spectra}
\end{figure}

In Fig.~\ref{fig:spectra}, we plot the spectrum of muon neutrinos and antineutrinos, after accounting for oscillations, as predicted in our model. The solid and dashed curves represent the contributions from pion decay and subsequent muon decay, respectively. Here, we have taken an injected spectral index of $\Gamma_p=2$ and adopted a high-scale cutoff, $E_{\rm max} \approx 60 \, {\rm PeV}$, as found using Eq.~\ref{eq:Emax} (for $B=10 \, {\rm MG}$). In calculating the synchrotron energy losses, we have accounted for the Poissonian nature of the pion and muon decays, integrating over the survival probability, $P =e^{-t/\gamma \tau}$. The black curves in this figure include the impact of synchrotron energy losses, adopting a magnetic field of $B=10 \, {\rm MG}$. The blue curves, in contrast, neglect the impact of synchrotron. We have normalized the spectrum to accommodate the IceCube data. Note that this normalization requires a luminosity in high-energy protons that is comparable to the X-ray luminosity of this source~\cite{Blanco:2023dfp}.

The spectrum of synchrotron radiation peaks at a value near the critical frequency, which is given by
\begin{align}
\nu_c &= \frac{3E^2 e B \sin \alpha_p}{4 \pi m^3} \\
&\approx 3 \, {\rm MeV} \times \bigg(\frac{E}{10 \, \rm TeV}\bigg)^2 \bigg(\frac{B}{10 \, {\rm MG}}\bigg) \bigg(\frac{m_{\pi}}{m}\bigg)^3 \sin \alpha_p \; . \nonumber
\end{align}
The synchrotron losses of very-high-energy pions and muons are thus expected to produce a significant flux of MeV-scale emission in this model, providing a promising target for future space-based gamma-ray detectors, such as AMEGO-X~\cite{AMEGO:2019gny} or e-ASTROGAM~\cite{e-ASTROGAM:2016bph}.

\section{Very Strong Magnetic Fields?}

As discussed in the previous section, the soft neutrino spectrum observed from NGC 1068 requires either that this source accelerates protons to a maximum energy of $E_{p}^{\rm max} \sim (2-6)\times 10^5 \, {\rm TeV}$, or that very strong magnetic fields are present within the acceleration region, $B \sim 10 \, {\rm MG}$. This is significantly larger than $B \sim \, {\rm kG}$ fields that have previously been considered in the context of this source~\cite{Blanco:2023dfp,Murase:2022dog,Das:2024vug}. Such extreme magnetic fields would have several important implications. First, they would substantially enhance synchrotron cooling of electrons, protons, charged pions, and muons. Second, magnetic fields of this order would push the system toward equipartition with the surrounding plasma, raising questions about their physical origin and stability.

The presence of a very large magnetic field in the corona of NGC 1068 would require a combination of a low value of the plasma beta parameter and/or a high value of the gas density. Plasma beta is defined as the ratio of the plasma pressure to the magnetic pressure, where the plasma pressure is the thermal pressure from the motion of the gas, and the magnetic pressure is related to the energy density of the magnetic field. Very low values (as small as $\beta \sim 10^{-6}$) have been reported in some disk environments~\cite{Hopkins:2023vjh}, but more recent work suggests a range of $\beta \sim 0.1-1$ for a magnetically dominated corona, as is believed to be the case for NGC 1068~\cite{Guo:2025glc,Fiorillo:2023dts}. This latter range would likely be unable to accommodate a magnetic field as strong as the one considered here. Alternatively, one could consider the role of the gas density. The magnetization parameter, $\sigma$, links the magnetic field strength directly to the gas density, $\sigma = B^2 / (4 \pi \rho_{\rm gas} c^2$) \cite{Fiorillo:2023dts}. For $\rho_{\rm gas} \, \sigma = 9 \times 10^{-6}$ kg/m$^3$ in an ion-dominated plasma, this naturally yields the magnetic field value adopted in this study~\cite{Mbarek:2023yeq}. In this framework, we conclude that magnetic fields of order $B \sim 10 \, {\rm MG}$ are more likely to be achieved through the presence of high gas densities than through low values of plasma beta.

\section{Predicted Flavor Ratios}

From the shape of the neutrino spectrum alone, it would be difficult to disentangle any potential impact of synchrotron cooling from the choice of the injected proton spectrum. Muon and pion decay, however, produce neutrinos in different ratios of flavors, providing us with a way to potentially distinguish between these two scenarios.

\begin{figure}[t]
\centering
\includegraphics[width=0.98\columnwidth]{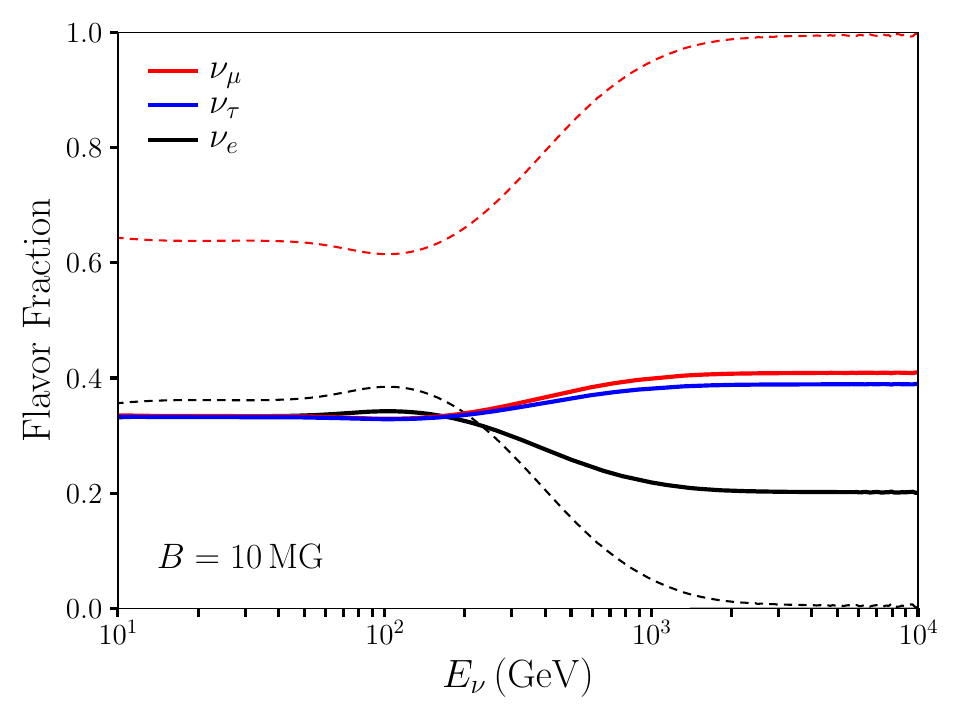}
\caption{The ratio of the neutrino flavors predicted from NGC 1068 in a model with an injected spectral index of \mbox{$\Gamma_p=2$,} a high-scale energy cutoff, $E^{\rm max}_p \approx 60 \,{\rm PeV}$, and a large magnetic field, $B=10 \, {\rm MG}$. The solid (dashed) lines take into account (neglect) the impact of standard neutrino oscillations.}
\label{fig:ratio}
\end{figure}

In Fig.~\ref{fig:ratio}, we plot the neutrino flavor ratios predicted from NGC 1068 for the same model and spectrum as shown in Fig.~\ref{fig:spectra}. Below $E_{\nu} \sim 200 \, {\rm GeV}$, synchrotron cooling is unimportant, and neutrinos are produced in the standard flavor ratios predicted from charged pion decay (including subsequent muon decay), $\nu_e:\nu_{\mu}:\nu_{\tau} = 1/3:2/3:0$. After accounting for oscillations, these ratios become
%
\begin{align}
\nu_e:\nu_{\mu}:\nu_{\tau} =\,\,\,\,\,\,\,\,\,\, &  \\ 
\frac{P_{e \rightarrow e} +   2 P_{{\mu} \rightarrow {e}}}{3} & :\frac{P_{e \rightarrow {\mu}}+2P_{{\mu} \rightarrow {\mu}}}{3}:
\frac{P_{e \rightarrow {\tau}}+2P_{{\mu} \rightarrow {\tau}}}{3} \; , \nonumber 
\end{align}
where $P_{i \rightarrow j}$ is the probability of oscillation from flavor $\nu_i$ to flavor $\nu_j$. For standard oscillation parameters, this reduces to 
\begin{align}
\nu_e:\nu_{\mu}:\nu_{\tau} \approx 0.32:0.34:0.34 \, .
\label{eq:ratio1}
\end{align}

At higher energies, at which muons undergo significant synchrotron losses before decaying, the resulting neutrinos will be purely of muon flavor. After oscillations, this becomes $\nu_e:\nu_{\mu}:\nu_{\tau} = P_{\nu_{\mu} \rightarrow \nu_{e}}:P_{\nu_{\mu} \rightarrow \nu_{\mu}}:P_{\nu_{\mu} \rightarrow \nu_{\tau}}$. Numerically, this results in
\begin{align}
\nu_e:\nu_{\mu}:\nu_{\tau} 
 \approx 0.20:0.41:0.39
\, .
\label{eq:ratio2}
\end{align}
The evolution with energy of the flavor ratios given in Eq.~\ref{eq:ratio1}, to those in Eq.~\ref{eq:ratio2}, is shown in Fig.~\ref{fig:ratio}.

\section{Neutrino Event Rates and Prospects for Future Measurement}

In the previous section, we showed that in the presence of a large magnetic field, muon cooling due to synchrotron can lead to changes in the neutrino flavor ratios from an astrophysical source. These changes, however, are relatively modest. To identify this effect would require flavor ratio measurements with a precision of approximately $\sim 20\%$.

The rate of neutrino-induced muon track events in a large-volume neutrino telescope can be written as
\begin{align}
\Gamma_{\mu} &= N_A A_{\rm eff}  \\
& \times \int \int_0^1 \frac{dN_{\nu_{\mu}}}{dE_{\nu_{\mu}}}(E_{\nu_{\mu}})  \, \frac{d\sigma^{\rm CC}_{\nu_{\mu} N}}{dy}(E_{\nu_{\mu}},y) \, R_{\mu}(E_{\mu}) \, dy  \, dE_{\nu_{\mu}} \; , \nonumber
\end{align}
%
where $N_A$ is the number density of nucleons in the detector medium, $A_{\rm eff}$ is the effective area of the detector, $d\sigma^{\rm CC}_{\nu_{\mu} N}/dy$ is the differential charged-current cross section with nucleons~\cite{Gandhi:1995tf}, and $y\equiv 1-(E_{\mu}/E_{\nu_{\mu}})$. The muon's range, $R_{\mu}$, is defined as the distance that a muon travels through the detector medium before falling below the energy threshold of the observatory. This can be expressed as
\begin{align}
R_{\mu} \approx \frac{1}{\beta} \ln\bigg(\frac{\alpha+E_{\mu}\beta}{\alpha+E_{\rm th} \beta}\bigg) \; ,
\end{align}
where $E_{\rm th}$ is the threshold energy of the detector, $\alpha \approx 0.002 \, {\rm GeV \, cm}^2/{\rm g}$, and $\beta\approx 6.2 \times 10^{-6} \, {\rm cm}^2/{\rm g}$~\cite{Seckel:2001pr}. For IceCube, we adopt $A_{\rm eff}=1 \, {\rm km}^2$ and $E_{\rm th} = 600 \, {\rm GeV}$. Similar parameters would apply to the case of the Cubic Kilometer Neutrino Telescope (KM3NeT)~\cite{KM3Net:2016zxf} or the Baikal Gigaton Volume Detector (GVD)~\cite{Baikal-GVD:2018isr}, which are currently under construction in the Mediterranean Sea and Lake Baikal, respectively.

The rate of neutrino-induced cascade events (also known as showers) can be estimated by
\begin{align}
\Gamma_{\rm cas} &= N_A V_{\rm eff}  \int_{E_{\rm th}^{\rm cas}}   \frac{dN_{\nu_{e}}}{dE_{\nu_{e}}}(E_{\nu_{e}})  \, \sigma^{\rm CC}_{\nu_{e} N}(E_{\nu_{e}})  \, dE_{\nu_{e}} \\
&+ N_A V_{\rm eff} \int_{E_{\rm th}^{\rm cas}/y} \int_0^1  \frac{dN_{\nu}}{dE_{\nu}}(E_{\nu})  \, \frac{d\sigma^{\rm NC}_{\nu N}}{dy}(E_{\nu},y) \, dy  \, dE_{\nu}, \nonumber
\end{align}
where $V_{\rm eff}$ is the effective volume of the detector and $E_{\rm th}^{\rm cas}$ is the threshold energy for electromagnetic or hadronic cascades. The first line in this expression corresponds to electromagnetic cascades produced in charged-current electron neutrino events, while the second line accounts for hadronic cascades from neutral-current events of all three neutrino flavors. The limits of integration have been chosen to select all cascades with a total energy above the stated threshold. For IceCube, we adopt $V_{\rm eff}=1 \, {\rm km}^3$ and $E_{\rm th}^{\rm cas} = 1 \, {\rm TeV}$.

The primary background in this case arises from atmospheric neutrinos. Using the spectrum of these particles as measured by Super-Kamiokande~\cite{Super-Kamiokande:2015qek}, the equations given above predict a rate of 19.6 muon tracks and 176.3 showers per year from the direction of NGC 1068 in a kilometer-scale telescope ($A_{\rm eff}=1 \, {\rm km}^2$, $V_{\rm eff}=1 \, {\rm km}^3$). In calculating these background rates, we have included muon tracks produced by atmospheric neutrinos from within a $1.5^{\circ}$ radius circle around NGC 1068. For showers, we consider a solid angle around NGC 1068 corresponding to the energy-dependent 50\% containment radius, as shown in Fig.~S5 of Ref.~\cite{IceCube:2023ame}, which varies from a radius of $\sim 17^{\circ}$ at $E_{\nu}=1 \, {\rm TeV}$ to $\sim 6^{\circ}$ at $E_{\nu}=1 \, {\rm PeV}$.

For the neutrino spectrum and flavor ratios shown in Figs.~\ref{fig:spectra} and~\ref{fig:ratio}, we calculate a rate of 8.61 muon tracks and 7.24 showers per year from NGC 1068 in a kilometer-scale telescope. In 8.7 years of observation (the duration of IceCube data considered in Ref.~\cite{IceCube:2022der}), this corresponds to $75.1\pm 15.7$ total muon tracks, and to a statistical significance of 4.8$\sigma$. This result is similar to the $70^{+22}_{-20}$ tracks ($4.2\sigma$) reported by the IceCube Collaboration. We further calculate a total number of $63.1\pm 40.4$ cascade events over this time, yielding an expected significance of 1.6$\sigma$.

If instead of the flavor ratios shown in Fig.~\ref{fig:ratio}, NGC 1068 produces neutrinos without significant muon or pion cooling (corresponding to the flavor ratios given in Eq.~\ref{eq:ratio1}), IceCube should see a a higher rate of cascades per muon track. Keeping the all-flavor neutrino spectrum fixed, we predict in this case a total number of $61.4\pm 15.3$ muon tracks and $75.9 \pm 40.2$ showers over 8.7 years of observation. This clearly cannot be distinguished from the event rates described in the previous paragraph with existing telescopes. To identify this signature of muon cooling on the neutrino flavor ratios from NGC 1068 would require a $\sim 30\%$ measurement of the shower rate. To make such a determination at $2\sigma$ ($5\sigma$) confidence would require IceCube to take data for $\sim 100$ ($\sim 600$) years. 

Fortunately, there are several proposals for high-energy neutrino telescopes with much larger volumes than those of IceCube, KM3NeT, or GVD. These future detectors include IceCube-Gen2~\cite{IceCube-Gen2:2020qha}, P-ONE~\cite{P-ONE:2020ljt}, TRIDENT~\cite{TRIDENT:2022hql}, and HUNT~\cite{Huang:2023mzt}. An instrument with a similar energy threshold and 10 km$^3$ of instrumented volume could measure the neutrino flavor ratios from NGC 1068 after 20 years of observation with enough precision to identify the effects of muon and pion cooling at approximately $3\sigma$ confidence.

\section{Summary and Conclusions}

The IceCube Neutrino Observatory has detected a statistically significant flux of TeV-scale neutrinos from the active galaxy NGC 1068. The neutrino spectrum of this source is very soft, falling rapidly with energy. This has generally been interpreted as an indication that NGC 1068 accelerates protons only up to a maximum energy of $E^{\rm max}_p \sim (2-6) \times 10^5 \, {\rm TeV}$. In this study, we have proposed an alternative scenario in which NGC 1068 accelerates protons to significantly higher energies, and the soft spectrum of the observed neutrinos is instead the result of the synchrotron energy losses that are experienced by charged pions and muons before they decay. 

The scenario proposed here would require a very strong magnetic field in the acceleration region of NGC 1068, on the order of $B \sim 10^7 \, {\rm G}$. If such a field is present, we estimate that this source could accelerate protons up to several tens of PeV, and that any charged pions or muons produced in the collisions of those protons would suffer significant energy losses due to synchrotron. In particular, muons and charged pions would cool in such an environment to energies of $E_{\mu} \sim 0.6 \, {\rm TeV}$ and $E_{\pi} \sim 10 \, {\rm TeV}$ before decaying, significantly softening of the neutrino spectrum in the energy range reported by IceCube.

If muons lose most of their energy through synchrotron before they can decay, this will impact the flavor ratios of the neutrinos produced by this source, increasing the observed ratio of track-like events to cascade-like events. Although existing observatories are not expected to be sensitive to the predicted magnitude of this effect, future telescopes with effective volumes on the order of $\sim 10 \, {\rm km}^3$ or greater could potentially use this information to identify the signatures of synchrotron cooling in NGC 1068.

\vspace*{0.4in}

\begin{acknowledgments}

We would like to thank Lorenzo Sironi, Alisa Galishnikova, and Kohta Murase for helpful discussions. DH is supported by the Office of the Vice Chancellor for Research at the University of Wisconsin-Madison, with funding from the Wisconsin Alumni Research
Foundation. TL is supported by the Swedish Research Council under contract 2022-04283. TL also acknowledges support from the Wenner-Gren Foundation under grant SSh2024-0037. We would like to thank the Mainz Institute for Theoretical Physics (MITP) of the Cluster of Excellence PRISMA+ (Project ID 390831469) where part of this work was carried out.

\end{acknowledgments}

\bibliography{ref}
\bibliographystyle{apsrev4-2}

\end{document}